\documentclass{article}
\begin{document}
\%Paper: astro-ph/9501049\\
\%From: FAIZULIN@univer.omsk.su\\
\%Date: Mon, 16 Jan 1995 14:44:22\\

 Let us consider first light stars numbering by  \cite{1}.
 There are some visual phenomena connected with the number by bright.

1. Greatest part of the stars (brighter then $2.4^m$)
  are concentrated at the straight lines which goes
 across stars with numbers $$2,3,5,6,7,10,11,15$$ for ($l,b$),
 ($\alpha,\delta$), ($\lambda,\beta$) coordinate systems.

2. It's so true for the group of $$2,6,7,10,15$$ for previous times:
 $-10^6,-2*10^6$ years.

3. If we consider  stars with numbers
 $$2,3,5,6,7,10,11,15,19,20,25,29,30,31,35,37,40,45,50$$
 then every of these stars concentreted at straight lines which goes
 across other two or more group stars.

4. If we consider group of stars with numbers
  $$ 3,4,6,7,8,9,10,11,13,17$$
 then greatest part of the nearest light galactic objects are
concentreted
 at the straight lines which goes across group stars.

5. There are connection between numbering and
  distances on the galactic planes ($l,b$).
 For example if we consider graph for group with numbers
 $$5,8,9,12,13,17$$
 then there are  proportionality between distances and numbers
 and  adjuncts to 20.

 Appendix: Numbers and coordinates of the 50 light stars.

  1 $\alpha Cma(1) $           6h 43   -16 -35;\\
  2 $\alpha Car(2) $          6h 23   -52 -40;\\
  3 $\alpha Bool(6)$            14h 13   19 27;\\
  4 $\alpha Lyr(4)$                    18h 35   38 44;\\
  5 $\alpha Cen(3) $            14h 36  -60 -38;\\
  6 $\alpha Aur(5)$             5h 13    45 57;\\
  7 $\beta Ori(7) $           5h 12    -8 -15;\\
  8 $\alpha Cmi(8)$                    7h 37     5 21;\\
  9 $\alpha Ori(12)$                    5h 52     7 24;\\
  10 $\alpha Eri(9)$                   1h 36   -57 -29;\\
  11$\beta Cen(10)$              14h 0   -60 -8;\\
  12 $\alpha Aql(11)$            19h 48    8 44;\\
  13 $\alpha Cru(13)$                   12h 24  -62 -49;\\
  14 $\alpha Tau(14)$                   4h 33    16 25;\\
  15$\alpha Sco(17)$                    16h 26  -26 -19;\\
  16 $\alpha Vir(16)$                   13h 23  -10 -54;\\
  17 $\beta Gem(15)$            7h 42    28 9;\\
  18 $\alpha Psa(18)$                   22h 55  -29 -53;\\
  19  $\beta Cru(21)$                   12h 45  -59 9;\\
  20 $\alpha Cyg(19)$             20h 40   45 6;\\
  21  $\alpha Leo(20)$         10h 6    12 13;\\
  22 $\epsilon Cma$                 6h 57   -28 54;\\
  23 $\alpha Gem$                   7h 31    32 0;\\
  24  $\lambda Sco $        17h 30  -37 -4;\\
  25 $\gamma Ori$                   5h 22     6 18;\\
  26 $\gamma Cru$                   12h 28  -56 50;\\
  27  $\beta Tau$                   5h 23    28 34;\\
  28 $\beta Car$                    9h 13   -69 -31;\\
  29 $\epsilon Ori$        5h 34    -1 -14;\\
  30 $\alpha Gru$                   22h 5   -47 -12;\\
  31 $\epsilon Uma$                 12h 52   56 14;\\
  32 $\zeta Ori$           5h 38    -1 -58;\\
  33 $\alpha Uma$                   11h 1    62 1;\\
  34 $\alpha Per$           3h 21    49 41;\\
  35 $\gamma Vel$                    8h 8    -47 -12;\\
  36 $\epsilon Sgr$        18h 21  -34 -25;\\
  37 $\delta CMa$                   7h 6    -26 -19;\\
  38 $\nu Uma$                      13h 46   49 34;\\
  39 $\epsilon Car$                 8h 22   -59 -21;\\
  40 $\theta Sco$           17h 34  -42 -58;\\
  41 $\beta Aur$                    5h 56    44 57;\\
  42 $\gamma Gem$            6h 35    16 27;\\
  43 $\alpha Tra$                   16h 43  -68 -56;\\
  44 $\delta Vel$                   8h 43   -54 -31;\\
  45 $\alpha Pav$                   20h 22  -56 -54;\\
  46 $\beta CMa$           6h 20   -17 -56;\\
  47 $\alpha Hya$                   9h 25    -8 -26;\\
  48 $o Cma$                        2h 17    -3 -12;\\
  49 $\alpha Ari$,                  2h 4     23 14;\\
  50 $\alpha UMi$           1h 49    89 2.

\end{document}